\documentclass{jfm}
\usepackage{graphicx}
\usepackage{epstopdf, epsfig}
\usepackage{color}
\usepackage{multirow}
\usepackage{natbib}
\usepackage{amsmath}
\usepackage[english]{babel}

\newtheorem{theorem}{Theorem}
\newtheorem{remark}{Remark}

\newcommand{\drawline}[2]{\raisebox{2.5pt}{\vbox{\hrule width #1 pt height #2 pt}}}
\newcommand{\spacce}[1]{\hspace{#1pt}}
\newcommand{\solid}{\nobreak\mbox{\drawline{24}{0.5}\spacce{2}}}

\newcommand{\ccline}{\drawline{5.5}{0.5}\spacce{2}\drawline{1}{0.5}\spacce{2}}
\newcommand{\dline}{\drawline{5.5}{0.5}\spacce{2}}
\newcommand{\dashdot}{\nobreak\mbox{\ccline\ccline\dline}}

\definecolor{orange}{rgb}{1,0.4,0}

\shorttitle{Near-wall turbulence intensity as $Re_\tau \rightarrow \infty$}
\shortauthor{Y. Hwang}

\title{Near-wall streamwise turbulence intensity \\ as $Re_\tau \rightarrow \infty$}

\author{Yongyun Hwang\corresp{\email{y.hwang@imperial.ac.uk}}}

\affiliation{Department of Aeronautics, Imperial College London,
South Kensington, SW7 2AZ, UK}

\begin{document}

\maketitle

\begin{abstract}
In this study, asymptotic scaling of near-wall streamwise turbulence intensity $\overline{u'u'}/u_\tau^2$ ($u_\tau$ is the friction velocity) is theoretically explored. The three scalings previously proposed are first reviewed with their derivation and physical justification: 1) $\overline{u'u'}/u_\tau^2 \sim \ln Re_\tau$ ($Re_\tau$ is the friction velocity; Marusic \& Kunkel, \emph{Phys. Fluids} vol. 15, no. 8, 2003, pp. 2461-2464); 2) $\overline{u'u'}/u_\tau^2 \sim 1/U_\infty^+$ ($U_\infty^+$ is the inner-scaled freestream velocity in boundary layer; Monkewitz \& Nagib, \emph{J. Fluid Mech.} vol. 783, 2015, pp. 474-503); 3) $\overline{u'u'}/u_\tau^2 \sim Re_\tau^{-1/4}$ (Chen \& Sreenivasan, \emph{J. Fluid Mech.} vol. 908, 2021, R3). A new analysis is subsequently developed based on velocity spectrum, and two possible scenarios are identified based on the asymptotic behaviour of the outer-scaling part of the near-wall velocity spectrum. In the former case, the outer-scaling part of the spectrum is assumed to reach a non-zero constant as $Re_\tau \rightarrow \infty$, and it results in the scaling of $\overline{u'u'}/u_\tau^2 \sim \ln Re_\tau$, both physically and theoretically consistent with the classical attached eddy model. In the latter case, a sufficiently rapid decay of the outer-scaling part of the spectrum with $Re_\tau$ is assumed due to the effect of viscosity, such that $\overline{u'u'}/u_\tau^2 < \infty$ for all $Re_\tau$. The following analysis yields $\overline{u'u'}/u_\tau^2 \sim 1/\ln Re_\tau$, asymptotically consistent with the scaling of $\overline{u'u'}/u_\tau^2 \sim 1/U_\infty^+$. The scalings are further verified with the existing simulation and experimental data and those from a quasilinear approximation (Holford \emph{et al.}, 2023, \texttt{arXiv:2305.15043}), the spectra of which all appear to favour $\overline{u'u'}/u_\tau^2 \sim 1/\ln Re_\tau$, although new datasets for $Re_\tau \gtrsim O(10^4)$ would be necessary to conclude this issue.
\end{abstract}

\section{Introduction}\label{sec:1}
\begin{figure} \vspace*{2mm}
\centering
\includegraphics[width=0.75\textwidth]{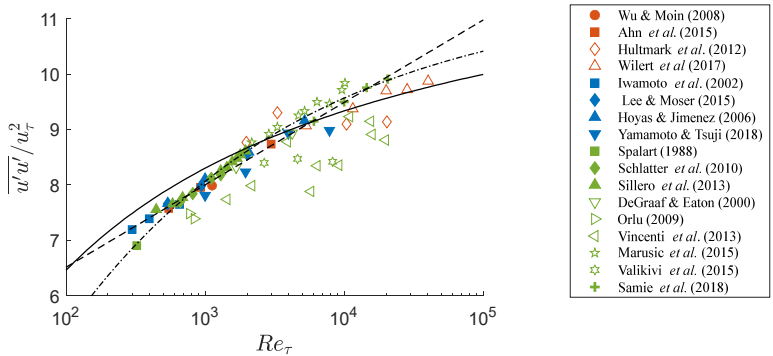}
\caption{Peak streamwise turbulence intensities from experiments (open symbols) and DNSs (filled symbols) and their scaling laws: \protect \dashed, $A_{S}+B_{S}\ln Re_\tau$ with $A_S=0.646$ and $B_S=3.54$ \cite[]{Samie2018}; \protect \solid, $A_{M}+B_{M}/U_{\infty}^+$ with $A_M=13$ and $B_M=100$ where the free stream velocity $U_{\infty}$ is given by $U_{\infty}^+=(1/\kappa)\ln Re_\tau+3.3$ with $\kappa=0.384$ \cite[]{Monkewitz2015,Nagib2022}; \protect \dashdot, $A_{C}+B_{C}\ln Re_\tau$ with $A_C=11.5$ and $B_C=19.32$ \cite[]{Chen2021}. The data from pipe, channel and boundary layer are denoted by the orange, blue and green colours, respectively. Here, $\overline{(\cdot)}$ indicates a time average and the superscript ${(\cdot)^+}$ denotes the inner-scaling with $u_\tau$ and $\nu$.}\label{fig1}
\end{figure}

The first-order moment (mean) of streamwise velocity of wall-bounded turbulence has a well-established layered structure in the wall-normal direction. In particular, in the near-wall region, the mean streamwise velocity is expressed in terms of a Reynolds-number-independent universal function, when normalised by the kinematic viscosity $\nu$ and the friction velocity $u_\tau$: i.e. the law of the wall \cite[][]{Prandtl25}. Early studies speculated that the second-order moment of the near-wall streamwise velocity would also behave in the same way \cite[e.g.][]{Sreenivasan1989,Mochizuki1996}. However, a large number of datasets, taken from accurate laboratory experiments and direct numerical simulations (DNS) since the 1990s, have revealed that the near-wall peak streamwise turbulence intensity consistently grows with Reynolds number. Figure \ref{fig1} reports the near-wall peak streamwise turbulence intensity in a range of the friction Reynolds number $Re_\tau$ from a number of previous experiments and DNSs \cite[e.g.][and many others]{Spalart1988,Klewicki1990,Osterlund1999,deGraaff2000,Metzger2001b,Marusic2001,Hoyas2006,Wu2008,Schlatter2009,Hultmark2012,Sillero2013,Vincenti2013,Vallikivi2015,Ahn2015,Lee2015,Willert2017,Samie2018}. Although the data are relatively scattered due to different flow geometry, experimental measurement methods, and numerical simulation methods, it is seen that the near-wall peak streamwise turbulence intensity grows slowly at least within the range of $Re_\tau$ considered.

The present study concerns the issue of how the growth of the peak intensity asymptotically scales as $Re_\tau \rightarrow \infty$. There are three different scaling laws that have been proposed so far: 1) $\ln Re_\tau$ scaling \cite[]{Marusic2003}; 2) Inverse of $U_{\infty}^+$ scaling \cite[$U_{\infty}$ is the freestream velocity in boundary layer and the superscript ${(\cdot)^+}$ denotes the inner-scaling with $u_\tau$ and $\nu$;][]{Monkewitz2015}; 3) $Re_\tau^{-1/4}$ scaling \cite[]{Chen2021}. As seen in figure \ref{fig1}, the three different scaling laws appear to be all reasonably good with the existing data, when the fitting constants are appropriately chosen, leading this issue to be unsettled without very accurate data available at higher $Re_\tau$. This paper will therefore begin first by reviewing these scaling laws with focus on their derivation, supporting evidence, limitation and physical relevance.

\subsection{$\ln Re_\tau$ scaling}\label{sec:1.1}
From their experimental data in a turbulent boundary layer, \cite{deGraaff2000} observed that $\overline{u'u'}/u_\tau^2$ slowly grows with $Re_\tau$ in a way that $\overline{u'u'}/u_\tau U_{\infty}$ remains approximately constant. Given that $U_{\infty}^+ \sim \ln Re_\tau$ at high $Re_\tau$, this also approximately implies the following scaling:
\begin{equation}\label{eq:1.1}
\frac{\overline{u'u'}}{u_\tau^2}\Big|_{y^+=y^+_{p}}=A_{S}+B_S\ln Re_\tau,
\end{equation}
where $y^+_{p}$ is the peak wall-normal location, and $A_{S}$ and $B_S$ are appropriate constants to be chosen. \cite{Marusic2003} pointed out that this observation is consistent with the classical attached eddy model \cite[]{Townsend1976,Perry1982,Perry1986}, when the near-wall streamwise velocity fluctuation is modelled to be
\begin{subequations}\label{eq:1.2}
\begin{equation}\label{eq:1.2a}
\frac{\overline{u'u'}(y^+)}{u_\tau^2}=f_1(y^+)+f_2(y/\delta),
\end{equation}
where
\begin{equation}\label{eq:1.2b}
f_2(y/\delta)=A_1-B_1 \ln \left(\frac{y}{\delta}\right)
\end{equation}
\end{subequations}
with $\delta$ being the outer length scale, such as half height of channel, radius of pipe and thickness of boundary layer. Here, $f_1(y^+)$ is a universal inner-scaling function which describes the streamwise turbulence intensity only from the viscous inner-scaling motions, and $f_2(y/\delta)$ depicts the contribution from the logarithmic and outer regions given by the classical attached eddy model \cite[]{Townsend1976}, where $B_1$ has often been referred to as the Townsend-Perry constant \cite[]{Marusic2013}. Note that if the peak location is assumed to be constant (say $y_{p}^+=15$), $\overline{u'u'}/{u_\tau^2}$ at the peak would be proportional to $\ln Re_\tau$ with $B_S=B_1 \ln y_p^+$ as in (\ref{eq:1.1}). Given that the Townsend-Perry constant proposed is $B_S\simeq 1.26$ \cite[e.g.][]{Marusic2013}, (\ref{eq:1.2}) would yield $B_S=3.41$, not very far from $B_S=3.54$ proposed by \cite{Samie2018} from their measurement of the peak intensity.

Perhaps, the scaling in (\ref{eq:1.1}) has been accepted most commonly \cite[see also e.g.][]{Smits2021}, as there is a considerable amount of evidence supporting the attached eddy hypothesis: i.e. the existence of the self-similar energy-containing motions, the size of which is proportional to the distance from the wall in the logarithmic region \cite[see also the recent review by][]{Marusic2019}. The attached eddy hypothesis itself has recently been demonstrated theoretically in terms of energetics \cite[]{Hwang2020b}, and there has been firm evidence on the existence of self-similar energy-containing eddies \cite[e.g.][]{Tomkins2003,delAlamo2006b,Hwang2011,Lozano-Duran2014,Hwang2015,Hwang2016,Hellstrom2016,Hwang2018,Baars2020a,Baars2020b} as well as supporting mathematical structure from the Navier-Stokes equations \cite[e.g.][]{Hwang2010b,Klewicki2013,Moarref2013,McKeon2017,Eckhardt2018,Yang2019,Doohan2019}.

The existence of the self-similar energy-containing motions in the logarithmic layer does not, however, necessarily imply that (\ref{eq:1.2}) is without any issues. Note that the expression (\ref{eq:1.2b}) is supposed to be valid in the logarithmic layer in the limit of $Re_\tau \rightarrow \infty$. It is also obtained by assuming that each of the self-similar energy-containing motions satisfies slip boundary condition (i.e. an inviscid theory), the key mathematical feature required to have the logarithmic term in (\ref{eq:1.2b}) \cite[]{Townsend1976}. However, any fluid motions in a viscous fluid must satisfy no-slip boundary condition, and the viscous effect is important in the near-wall region, where the scaling of interest concerns. In fact, \cite{Hwang2016c} showed that there is non-trivial Reynolds-number dependent viscous effect on the spectra of the footprint of self-similar energy-containing motions (i.e. the near-wall part of the motions). Even in the entire logarithmic layer, the viscous effect is not negligible as long as $Re_\tau$ is finite. In the upper logarithmic layer (or inertial sublayer), it was recently shown that both $A_1$ and $B_1$ must vary with $Re_\tau$ theoretically \cite[]{Hwang2022}. In the lower logarithmic layer \cite[or mesolayer; see also][]{Afzal1982,Klewicki2013}, an expression different from (\ref{eq:1.2b}) is needed to incorporate the viscous effect: for example, see (4.6) in \cite{Hwang2022}.

\subsection{Inverse of $U_{\infty}^+$ scaling}\label{sec:1.2}
\cite{Monkewitz2015} proposed a scaling of near-wall peak streamwise turbulence with $Re_\tau$ different from (\ref{eq:1.1}). The proposed scaling reads as
\begin{equation}\label{eq:1.3}
\frac{\overline{u'u'}}{u_\tau^2}\Big|_{y^+=y^+_{p}}=A_{M}+\frac{B_M}{U_{\infty}^+}+O\left(\frac{1}{{U_{\infty}^+}^2}\right),
\end{equation}
where $A_M$ and $B_M$ are constants. 
The scaling (\ref{eq:1.3}) is obtained by an asymptotic analysis of streamwise mean momentum equation of turbulent boundary layer combining with the existing DNS data. Using the von K\'{a}rm\'{a}n integrated momentum equation, \cite{Monkewitz2015} showed that the small parameter required for the asymptotic balance is $1/U_\infty^+$. The resulting ${\overline{u'u'}}/u_\tau^2$ is then finite at $Re_\tau \rightarrow \infty$ and proportional to $1/U_\infty^+$ for sufficiently high $Re_\tau$. 

The scaling (\ref{eq:1.3}) provides a good fit for the existing dataset (solid line in figure \ref{fig1}) \cite[see also][]{Monkewitz2022}. It is also probably the most rigorous one among the three scalings available, given that it is obtained by a direct analysis of the Navier-Stokes equations using DNS data. Perhaps, the only limitation is that it is strictly applicable to boundary layer (i.e. spatially developing flow), where the mean momentum equation contains non-vanishing ${\overline{u'u'}}$: for example, ${\overline{u'u'}}$ does not appear in the streamwise mean momentum equation in internal parallel shear flows, such as channel and pipe, thereby not being able to straightforwardly justify (\ref{eq:1.3}) for such flows. Nevertheless, recent studies, which applied quasilinear approximations to the Navier-Stokes equations in channel flow \cite[]{Hwang2020a,Skouloudis2021,Holford2023b}, have repeatedly suggested that their peak streamwise turbulence intensity consistently follows the scaling in (\ref{eq:1.3}).

\subsection{$Re_\tau^{-1/4}$ scaling}\label{sec:1.3}
The final scaling of interest in this study has recently been proposed by \cite{Chen2021}:
\begin{equation}\label{eq:1.4}
\frac{\overline{u'u'}}{u_\tau^2}\Big|_{y^+=y^+_{p}}=A_{C}+B_C Re_\tau^{-1/4}+O({y^+}^3),
\end{equation}
where $A_C$ and $B_C$ are constants to be determined with the measured data. The starting point of (\ref{eq:1.4}) is from the following equation, obtained by applying a Taylor expansion about the wall \cite[see also][for all Reynolds stress components]{Smits2021}:
\begin{equation}\label{eq:1.5}
\frac{\overline{u'u'}(y)}{u_\tau^2}=\epsilon_W^+{y^+}^2 + O({y^+}^3),
\end{equation}
where $\epsilon_W^+$ is inner-scaled turbulence dissipation at the wall. Two key conjectures are subsequently made to estimate the scaling behavior of ${\overline{u'u'}}/u_\tau^2$ in the near-wall region. First, from the well-known fact that the inner-scaled turbulence production is bounded by $1/4$ as $Re_\tau \rightarrow \infty$, \cite{Chen2021} argued that $\epsilon_W^+$ is also expected to be bounded by $\epsilon_\infty^+(=1/4)$. The defect dissipation from this bound, $\epsilon_d$, may be defined as
\begin{subequations}
\begin{equation}\label{eq:1.6a}
\epsilon_d^+=1/4-\epsilon_W^+.
\end{equation}
Second, they argued that the defect dissipation may be related to bursting in the near-wall region \cite[e.g.][]{Kline1967}, which transports some of the turbulent kinetic energy to the outer region without being dissipated locally. The time scale of this process is subsequently hypothesized to be $\eta_o/u_\tau$, where $\eta_o$ is the Kolmogorov length scale in the outer region: i.e. $\eta_0=\nu^{3/4}/\epsilon_o^{1/4}$, where $\epsilon_o=u_\tau^3/\delta$. If this is so, $\epsilon_d^+ \sim Re^{-1/4}$. Combining with (\ref{eq:1.6a}), this leads to
\begin{equation}\label{eq:1.6b}
\epsilon_W^+=1/4-\beta Re_\tau^{-1/4},
\end{equation}
\end{subequations}
where $\beta$ is a constant to be determined.


The scaling (\ref{eq:1.5}) provides a good fit for the existing dataset (dash-dotted line in figure \ref{fig1}). However, apart from the influence of the higher-order term in (\ref{eq:1.5}), unfortunately, the two conjectures made to derive (\ref{eq:1.5}) in \cite{Chen2021} do not have supporting evidence. First, there is no physical reason for $\epsilon_W^+$ to be bounded by the peak near-wall production, 1/4. Wall-bounded turbulence is largely non-local in the wall-normal direction, and there is a large amount of evidence that the near-wall region is subject to influence of the structures originating from the logarithmic and the outer regions \cite[see][and many others]{Hutchins2007,Mathis2009,Mathis2013,Duvvuri2015,Agostini2016,Zhang2016}. Importantly, the turbulent energy transport caused by the resulting near-wall inner-outer interaction is directly balanced with dissipation in the near-wall region \cite[]{Cho2018}. In other words, unlike the production only affected by inner scale \cite[see the Reynolds shear stress spectra (figure 10$e$) in][]{Hwang2016c}, dissipation in the near-wall region is increasingly influenced by the wall-attached part of the energy-containing motions from the log- and outer regions upon increasing $Re_\tau$. This implies that $\epsilon_W^+$ needs not to be bounded by the peak production value 1/4 in the limit of $Re_\tau\rightarrow \infty$ -- it is possible for the near-wall region to dissipate out the near-wall part of structures associated with infinitely many scales varying from $\delta_\nu(\equiv \nu/u_\tau)$ to $\delta$, given that $Re_\tau =\delta/\delta_{\nu}$ by definition. In fact, we shall see in \S\ref{sec:2} that the increase of the near-wall peak intensity with $Re_\tau$ is because there is an increasing contribution of the wall-attached part of the motions originating from the log- and outer regions (figure \ref{fig3}), a physical picture rather consistent with attached eddy hypothesis of \cite{Townsend1976}. 

Second, there is no evidence that bursting transports the near-wall turbulence all the way up to the outer region especially for $Re_\tau \rightarrow \infty$. In fact, it is quite possible that the near-wall turbulence transported upwards by bursting may well be dissipated in the logarithmic region, where most of dissipation should take place at high $Re_\tau$ \cite[]{Tennekes1967}: see also data from \cite{Lee2015,Lee2019}. Let us assume that the transport process takes place at the time scale of $\eta_o/u_\tau$, as conjectured by \cite{Chen2021}. Given that $v \sim O(u_\tau)$ throughout the entire wall-normal location ($v$ is the wall-normal velocity), the wall-normal turbulent transport (or the wall-normal advective flux for turbulent fluctuation) can take place only with the speed of $O(u_\tau)$. The distance that the near-wall turbulence can travel upwards is then only $O(\eta_o)$ over the time scale of $O(\eta_o/u_\tau)$. Given $\eta_o/\delta=Re_\tau^{-3/4}$, it is evident that the near-wall turbulence is increasingly unable to reach the outer region as $Re_\tau \rightarrow \infty$. It will remain at best in the lower logarithmic region (or mesolayer), given that $\delta_m/\delta \sim Re_\tau^{-1/2}$ \cite[$\delta_m$ is the upper boundary of the mesolayer; see][]{Afzal1982,Sreenivasan1997,Wei2005}. This suggests that the second conjecture of \cite{Chen2021} is self-contradictory. More importantly, in the spanwise minimal channel \cite[e.g.][]{Hwang2013}, where the bursting is allowed to transport near-wall turbulence from the wall in the absence of the motions from the logarithmic and outer regions, the near-wall streamwise turbulence intensity does not change with $Re_\tau$, not supporting the second conjecture of \cite{Chen2021}.

\subsection{Scope of the present study}\label{sec:1.4}

Thus far, the relevance of the three scalings of the near-wall peak streamwise turbulence intensity has been reviewed. The scaling in (\ref{eq:1.1}) has a good physical grounds as a large number of recent investigations have confirmed the existence of statistically and dynamically self-similar energy-containing motions in the logarithmic layer (i.e. attached eddies). However, it remains unclear if the simple extension of the classical attached eddy model described in (\ref{eq:1.2}) would accurately incorporate the near-wall viscous effect into streamnwise turbulence intensity. The scaling in (\ref{eq:1.3}) is perhaps the most rigorous at least for boundary layer, as (\ref{eq:1.3}) is directly derived from a self-consistent asymptotic analysis of the Naiver-Stokes equations \cite[]{Monkewitz2015}. Nevertheless, the physical implication of this scaling is unclear especially in relation to the dynamics of the flow (e.g. coherent structures, energy balance, etc). Lastly, the scaling (\ref{eq:1.5}) does not appear to have strong supporting physical evidence, allowing for some critical counter arguments with a room for a further debate.

\begin{figure} \vspace*{2mm}
\centering
\includegraphics[width=0.95\textwidth]{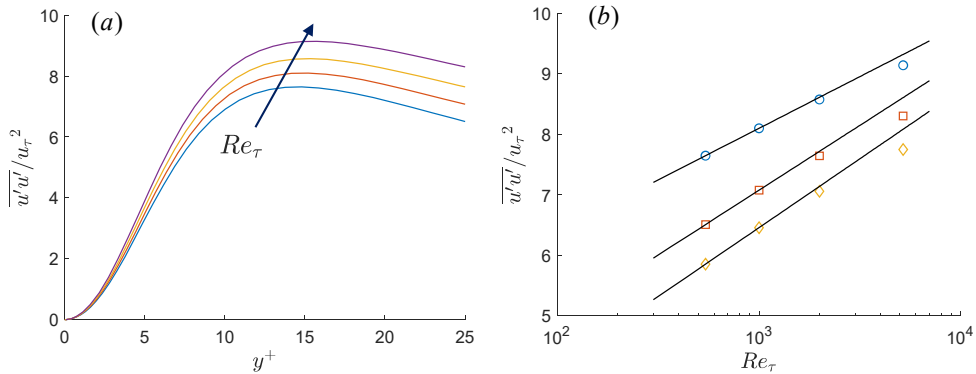}
\caption{Near-wall streamwise turbulence intensity in channel for $Re_\tau=544, 1000, 1995, 5186$ \cite[]{Lee2015}: ($a$) near-wall profile of $\overline{u'u'}(y^+)/u_\tau^2$; $(b)$ values of $\overline{u'u'}(y^+)/u_\tau^2$ at $y^+=y^+_{\mathrm{peak}}, 25,30$ with $y^+_\mathrm{peak}\simeq 15$. In $(b)$, the solid black lines indicate fits in the form of $A \log Re_\tau+B$, where $A$ and $B$ are determined by the data at two lowest Reynolds numbers.}\label{fig2}
\end{figure}

Given the discussion above on the three different scaling laws, the objective of the present study is to theoretically explore the possible asymptotic scaling behaviours of near-wall streamwise turbulence intensity as $Re_\tau \rightarrow \infty$. In \S\ref{sec:2}, some observations will first be made on the near-wall streamwise velocity and corresponding spectrum with DNS \cite[]{Lee2015} and experimental \cite[]{Samie2018} data. Based on this observation, an exact form of the near-wall streamwise turbulence intensity is formulated using a dimensional analysis of the corresponding velocity spectra. In \S\ref{sec:3}, it will be seen that the near-wall spectrum for $k_x\delta \sim O(1)$ ($k_x$ is the streamwise wavenumber) has a crucial importance in the prediction of scaling. Depending on its behaviour, two scaling behaviours will be shown to be possible: 1) ${\overline{u'u'}}/u_\tau^2 \sim \ln Re_\tau$; 2) ${\overline{u'u'}}/u_\tau^2 \sim (\ln Re_\tau)^{-1}$. The near-wall spectra from the existing DNS \cite[]{Lee2015} and experimental data \cite[]{Samie2018} will be seen to favour the latter case at least within the range of $Re_\tau$ currently available. In \S\ref{sec:4}, the relevance of the latter scaling will be discussed further using the data generated by a quasilinear approximation of the Navier-Stokes equations for channel flow up to $Re_\tau=10^5$ \cite[]{Holford2023b}. This work will be concluded with some remarks in \S\ref{sec:5}.

\section{Problem formulation}\label{sec:2}

\subsection{Some observations on near-wallturbulence intensity}\label{sec:2.1}
Before exploring the possible scaling behaviour, it would be useful to start by making some observations on the features of near-wall turbulence intensity. Figure \ref{fig2} shows the behaviour of near-wall streamwise turbulence intensity in channel flow from \cite{Lee2015} on increasing $Re_\tau$. The growth of the near-wall turbulence intensity is evident over the entire near-wall region, as seen in figure \ref{fig2}($a$). The peak wall-normal location appears at $y^+_\mathrm{peak}\simeq 15$, but there is also a very weak dependence on $Re_\tau$, the feature also pointed out by some previous studies \cite[e.g.][]{Willert2017,Hwang2020a}. Note that all the three scalings discussed in \S\ref{sec:1} assume the inner-scaled peak location is not a function of $Re_\tau$. Although the derivation of (\ref{eq:1.3}) could suitably be modified without change in its form (see \S\ref{sec:3.2}), figure \ref{fig2}($a$) implies that, in principle, comparing the peak streamwise turbulence intensity from DNS/experimental data with all the three scalings discussed in \S\ref{sec:1} is not precisely valid, especially when a wide range of $Re_\tau$ encompassing several decades is to be considered. Instead, the streamwise turbulence intensity at a fixed inner-scaled wall-normal location must be considered to be directly compared with the scalings given in \S\ref{sec:1}. Figure \ref{fig2}($b$) shows the scaling behaviour of near-wall turbulence intensity at a few more locations in the near-wall region. It becomes more evident even at relatively low $Re_\tau (\lesssim 5200)$ that the near-wall turbulence intensity deviates from (\ref{eq:1.1}), indicating that the DNS data favours the alternative scalings, (\ref{eq:1.3}) or (\ref{eq:1.4}).

\begin{figure} \vspace*{2mm}
\centering
\includegraphics[width=0.95\textwidth]{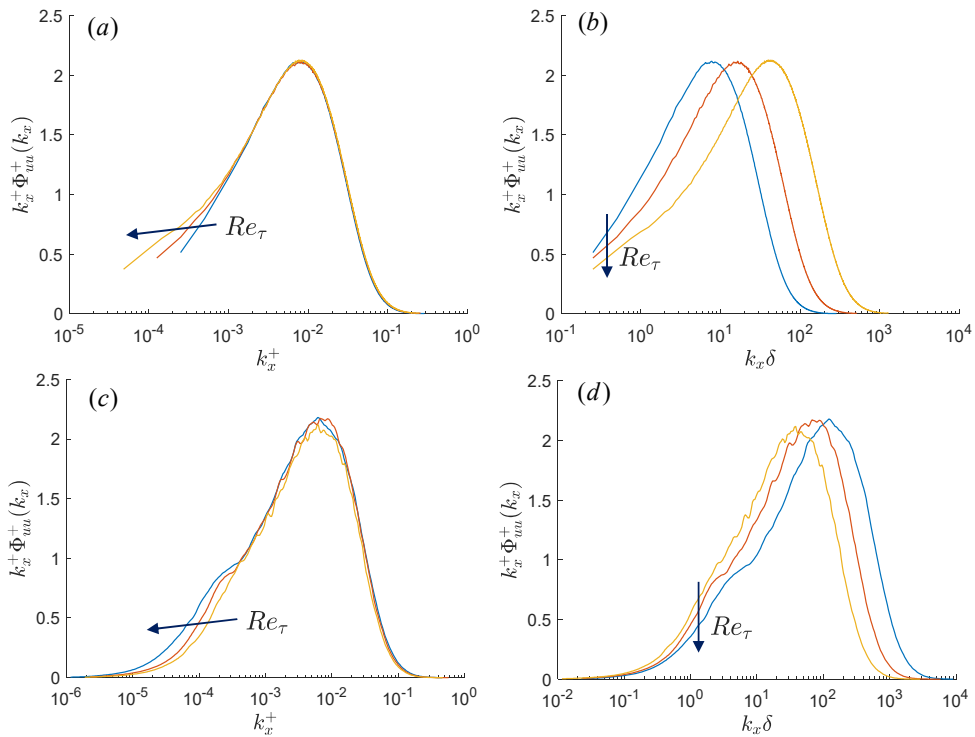}
\caption{Premultiplied streamwise wavenumber spectra of streamwise velocity at $y^+=15$ in $(a,b)$ channel \cite[DNS from][]{Lee2015} and $(c,d)$ boundary layer \cite[experiment from][]{Samie2018}: $(a,c)$ inner scaling; $(b,d)$ outer scaling. In $(a,b)$, $Re_\tau=1000, 1995, 5186$, and, in $(c,d)$, $Re_\tau=6123, 10100, 19680$. }\label{fig3}
\end{figure}

To understand the behaviour of near-wall streamwise turbulence intensity in figure \ref{fig2} better, we consider its power spectral density with respect to the streamwise wavenumber $k_x$, defined with
\begin{equation}\label{eq:2.1}
\overline{u'u'}(y)=\int_0^\infty \Phi_{uu}(k_x,y) \mathrm{d}k_x=\int_{-\infty}^{\infty} k_x \Phi_{uu}(k_x,y) \mathrm{d}(\ln k_x).
\end{equation}
Figure \ref{fig3} visualises pre-multiplied power spectral density of streamwise velocity at $y^+=15$ with respect to inner- and outer-scaled streamwise wavenumber. Here, the data are from \cite{Lee2015} for channel (figures \ref{fig3}$a,b$) and from \cite{Samie2018} for boundary layer (figures \ref{fig3}$c,d$). When each spectrum from different $Re_\tau$ is scaled with inner units (figures \ref{fig3}$a,c$), all of them collapse into a single curve with the maximum intensity of $k_x^+ \Phi_{uu}^+(k_x^+)\simeq 2.2$ for $k_x^+ \gtrsim O(10^{-3})$ ($\lambda_x^+ \lesssim O(10^{3})$ where $\lambda_x=2\pi/k_x$). In particular, the peak occurs at $k_x^+ \simeq 0.007$ ($\lambda_x^+ \simeq 1000$), close to the streamwise length of near-wall streaks \cite[]{Hwang2013}. As $Re_\tau$ is increased, the spectrum gradually extends to smaller $k_x^+$, forming a long tail at $k_x^+ \lesssim O(10^{-3})$ for $Re_\tau$ considered here. The presence of the spectral tail for small $k_x$ (or large $\lambda_x$) is evidently the reason why the peak streamwise turbulence intensity grows with $Re_\tau$.

Given that the largest length scale of the flow is $\delta$, the spectra with respect to the outer-scaled streamwise wavenumber are further plotted in figure \ref{fig3}$(c,d)$. Both DNS and experimental data show that there is a considerable amount of spectral energy at $k_x\delta \lesssim 1$ ($\lambda_x/\delta \gtrsim 2\pi$). As discussed in \S\ref{sec:1}, this is associated with the penetration of very large-scale motions (or $\delta$-scaling long streaky motions) into the near-wall region \cite[e.g.][]{Hutchins2007,Mathis2009,Mathis2013,Duvvuri2015,Agostini2016,Zhang2016}. Note that the spectral energy at this wavenumber range decays with $Re_\tau$, suggesting that the near-wall influence (or footprint) of $\delta$-scaling long streaky motions appears to diminish. This is the non-trivial viscous wall effect on the footprint of $\delta$-scaling structures \cite[]{Hwang2016c}, and its origin currently remains not understood. Assuming a gradual diminishment of energy for $k_x\delta \lesssim 1$ on increasing $Re_\tau$, most of the spectral energy related to the growth of $\overline{u'u'}/u_\tau^2$ with $Re_\tau$ is expected to originate from $O(1/\delta)\lesssim k_x \lesssim O(1/\delta_\nu)$, as $Re_\tau \rightarrow \infty$. By definition, these streamwise wavenumbers (or length scales) are associated with the logarithmic layer (i.e. the overlap region). Indeed, the near-wall spectral energy for $O(1/\delta)\lesssim k_x\lesssim O(1/\delta_\nu)$ has consistently been understood to be the contribution of `inactive' motions of self-similar attached eddies of \cite{Townsend1976} \cite[for a detailed discussion, see also][]{Hwang2015,Hwang2016c,Deshpande2020,Holford2023a}. It is worth mentioning that the logarithmic term in the scaling of (\ref{eq:1.1}) theoretically originates from such inactive motions, the consequence of allowing for wall-parallel fluid motions near the wall through the slip boundary condition \cite[]{Townsend1976}. In this respect, the scaling in (\ref{eq:1.1}) is still sound with DNS and experimental data at least from a physical viewpoint, although the poorly understood viscous effect presumably undermines its precise relevance to viscous fluids. 

\subsection{Near-wall turbulence intensity}\label{sec:2.2}
\begin{figure} \vspace*{2mm}
\centering
\includegraphics[width=0.75\textwidth]{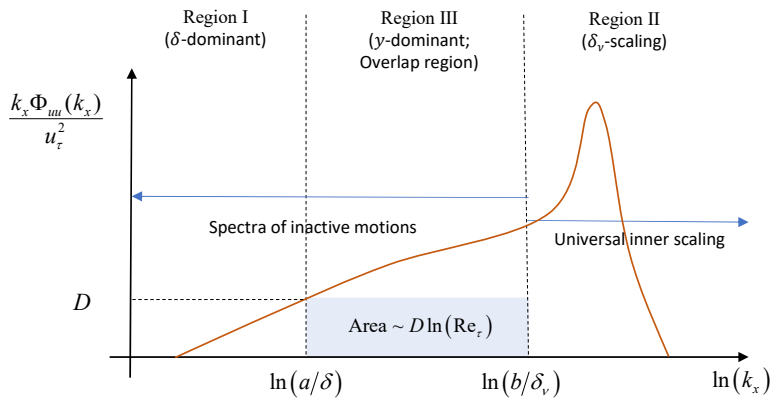}
\caption{A schematic diagram of the premultiplied one-dimensional spectrum of near-wall streamwise velocity at a fixed inner-scaled wall-normal location ($y^+=c$).}\label{fig4}
\end{figure}

Given the discussion in \S\ref{sec:2.1}, the behaviour of the streamwise velocity spectrum in the near-wall region is crucial to understand the peak scaling of $\overline{u'u'}/u_\tau^2$ with $Re_\tau$. Here, a further analysis will be presented for the streamwise wavenumber velocity spectrum, a function of $u_\tau$, $\nu$ and $\delta$, $k_x$ and $y$. A schematic diagram of the spectrum is first considered at a fixed inner-scaled wall-normal location (i.e. $y^+=c$) close to the peak location of $\overline{u'u'}/u_\tau^2$, as in figure \ref{fig4}. The structure of spectrum is set to be divided into three regions, depending on the value of $k_x$: 1) Region I, where $k_x \lesssim O(1/\delta)$; 2) Region II, where $k_x \gtrsim O(1/\delta_\nu)$; 3) Region III, where $O(1/\delta) \lesssim k_x \lesssim O(1/\delta_\nu)$.

Region I is defined to be $k_x \delta \leq a$ ($a$ is an appropriate constant to be chosen from data) at a fixed inner-scaled wall-normal location (say $y^+=c$), leading to $k_x \sim  O(1/\delta)$ and $y \sim O(\delta_\nu)$. Combining this with the Buckingham $\mathrm{\Pi}$ theorem and (\ref{eq:2.1}), the spectrum is given by
\begin{subequations}\label{eq:2.2}
\begin{equation}\label{eq:2.2a}
\frac{\Phi_{uu}(k_x,y;\delta,\delta_\nu)}{u_\tau^2}=\frac{\delta\Phi_{uu}(k_x \delta,y^+; Re_\tau)}{u_\tau^2}=\delta g_1(k_x \delta, y^+; Re_\tau).
\end{equation}
At $y^+=c$, the corresponding dimensionless pre-multiplied spectrum is then written as
\begin{equation}\label{eq:2.2b}
\frac{k_x\Phi_{uu}(k_x;\delta,\delta_\nu)}{u_\tau^2}=\frac{k_x\delta\Phi_{uu}(k_x \delta; Re_\tau)}{u_\tau^2}=k_x\delta g_1(k_x \delta; Re_\tau)=h_1(k_x \delta; Re_\tau).
\end{equation}
\end{subequations}
Now, it becomes evident that the dimensionless pre-multiplied spectrum for $k_x\delta \sim O(1)$ must be a function of $Re_\tau$ in general, consistent with the DNS and experimental data shown in figure \ref{fig3}$(b,d)$.

Similarly to Region I, Region II is defined at $y^+=c$ for $k_x^+ \geq b$, where $b$ is an appropriate constant, yielding $k_x \sim  O(1/\delta_\nu)$ and $y \sim O(\delta_\nu)$. It is therefore expected that $\delta$ would no longer be relevant in Region II. It is also worth noting that $\delta_\nu$ is the Kolmogorov length scale in the near-wall region, as the dissipation rate is given by $\epsilon=u_\tau^3/\delta_\nu$. This implies that the smallest possible length scale of the spectrum is $\delta_\nu$ and that turbulence production and dissipation takes place at the same length scale in the near-wall region, like in typical transitional flows. Therefore, it is expected that there is no spectrum developed for energy cascade (i.e. the $k_x^{-5/3}$ law) in Region II. The Buckingham $\mathrm{\Pi}$ theorem and (\ref{eq:2.1}) subsequently lead the spectrum to take the following form:
\begin{subequations}\label{eq:2.3}
\begin{equation}
\frac{\Phi_{uu}(k_x,y;\delta_\nu)}{u_\tau^2}=\frac{\delta_\nu\Phi_{uu}(k_x^+,y^+)}{u_\tau^2}=\delta g_{2}(k_x \delta, y^+).
\end{equation}
The corresponding dimensionless pre-multiplied spectrum at $y^+=c$ is given by
\begin{equation}
\frac{k_x\Phi_{uu}(k_x)}{u_\tau^2}=\frac{k_x^+\Phi_{uu}(k_x^+)}{u_\tau^2}=k_x^+ g_{2}(k_x^+)=h_{2}(k_x^+).
\end{equation}
\end{subequations}
Unlike (\ref{eq:2.2b}), the spectrum in Region II does not have any dependence on $Re_\tau$, since $\delta$ is no longer a relevant length scale here. This is consistent with the DNS and experimental data in figure \ref{fig3}$(a,c)$, where the spectra from different $Re_\tau$ are collapsed into a single curve.

Region III is defined for $a/\delta \leq k_x  \leq b/\delta_\nu$. If $h_1(k_x \delta; Re_\tau)$ and $h_{2}(k_x^+)$ are known, an asymptotic matching may be proceeded to understand the form of the spectrum for $O(1/\delta) \ll k_x \ll O(1/\delta_\nu)$ as in \cite{Perry1986} for the logarithmic layer. It will be seen in \S\ref{sec:3.1} that such an analytical progress is possible for a particular case without knowing the full details of $h_1(k_x \delta; Re_\tau)$ and $h_{2}(k_x^+)$. However, in general, the dimensionless pre-multiplied spectrum in Region III is expected to be a function of $Re_\tau$, given that the spectrum in Region I is a function of $Re_\tau$: i.e.
\begin{equation}\label{eq:2.4}
\frac{k_x\Phi_{uu}(k_x)}{u_\tau^2}=h_{3}(k_x l; Re_\tau),
\end{equation}
where $l$ is an appropriate length scale given by $\delta_\nu < l < \delta$.

Using (\ref{eq:2.1}) and the features of the spectra from (\ref{eq:2.2}) to (\ref{eq:2.4}), the streamwise turbulence intensity at $y^+=c$ may be written as follows:
\begin{subequations}\label{eq:2.5}
\begin{eqnarray}\label{eq:2.5a}
\frac{\overline{u'u'}}{u_\tau^2}\Big|_{y^+=c}&&=\underbrace{\int_{-\infty}^{\ln (a/\delta)} \frac{k_x \Phi_{uu}(k_x,c\delta_\nu)}{u_\tau^2} \mathrm{d}(\ln k_x)}_{\mathrm{Region~I}}+\underbrace{\int_{\ln (b/\delta_\nu)}^{\infty} k_x \Phi_{uu}(k_x,c\delta_\nu) \mathrm{d}(\ln k_x)}_{\mathrm{Region~II}}\nonumber \\
&& +\underbrace{\int_{\ln (a/\delta)}^{\ln (b/\delta_\nu)} \frac{k_x \Phi_{uu}(k_x,c\delta_\nu)}{u_\tau^2} \mathrm{d}(\ln k_x)}_{\mathrm{Region~III}} \nonumber \\
&& = A(Re_\tau)+C+B(Re_\tau)\left[\ln Re_\tau+\ln \left(\frac{b}{a}\right)\right],
\end{eqnarray}
where
\begin{equation}\label{eq:2.5b}
A(Re_\tau)=\int_{-\infty}^{\ln a} h_1(k_x \delta; Re_\tau) \mathrm{d}(\ln (k_x \delta))
\end{equation}
from Region I,
\begin{equation}\label{eq:2.5c}
C=\int_{\ln b}^{\infty} h_{2}(k_x^+) \mathrm{d}(\ln k_x^+)
\end{equation}
from Region II, and
\begin{equation}\label{eq:2.5d}
B(Re_\tau)=\left[\ln \left(\frac{bl}{\delta_\nu}\right)-\ln \left(\frac{al}{\delta}\right)\right]^{-1} \int_{\ln (al/\delta)}^{\ln (bl/\delta_\nu)} h_3(k_x l; Re_\tau) \mathrm{d}(\ln (k_x l))
\end{equation}
from Region III. Here, the value of $B(Re_\tau)$ must be comparable to a typical value of dimensionless premultiplied power-spectral intensity for $a/\delta \leq k_x  \leq b/\delta_\nu$, because the mean value theorem for integral indicates the existence of $B(Re_\tau)$, satisfying
\begin{equation}\label{eq:2.5e}
B(Re_\tau)=h_{3}(k_x^* l; Re_\tau),
\end{equation}
where $k_x^*$ is a streamwise wavenumber given in the range of $a/\delta \leq k_x^*\leq b/\delta_\nu$.
\end{subequations}

It is also worth noting that $a$ and $b$ are appropriate constants to be chosen. While the choice for $b$ must ensure $C$ in (\ref{eq:2.5c}) to be a constant, the choice for $a$ is rather arbitrary as long as $a \lesssim O(1)$. In fact, choosing a sufficiently small $a(\ll 1)$ can further simplify (\ref{eq:2.5}) into
\begin{subequations}\label{eq:2.6}
\begin{equation}\label{eq:2.6a}
\frac{\overline{u'u'}}{u_\tau^2}\Big|_{y^+=c}=C+B(Re_\tau)\left[\ln Re_\tau+\ln \left(\frac{b}{a}\right)\right]+O(a^2),
\end{equation}
where the contribution from Region I (i.e. $A(Re_\tau)$) is estimated to be
\begin{equation}
A(Re_\tau)=\int_{0}^{a/\delta} \frac{\Phi_{uu}(k_x)}{u_\tau^2} \mathrm{d}(k_x)=\int_{0}^{a} \frac{\Phi_{uu}(k_x \delta)}{\delta u_\tau^2} \mathrm{d}(k_x \delta)\sim O(a^2)
\end{equation}
\end{subequations}
from $\Phi_{uu}(k_x\delta) \sim k_x\delta$ for $k_x\delta \ll 1$. For example, if $a=0.01$ is chosen for the experimental data of \cite{Samie2018}, $A(Re_\tau) \sim O(10^{-4})$ and this is seen to be negligibly small (figure \ref{fig3}$d$). However, such a small choice of $a$ does not lead to any singular behaviour of $\ln (b/a)$ in (\ref{eq:2.6a}) because of its logarithm. Indeed, if $b=10^{-3}$ is chosen with $a=0.01$ from the data in figures \ref{fig3}$(a,c)$, $\ln (b/a) \sim O(1)$.

\section{Two possible scenarios}\label{sec:3}
The formulation in \S\ref{sec:2.2} suggests that the increase of $\overline{u'u'}/u_\tau^2$ in the near-wall region is due to the increasing contribution of the motions, the streamwise length scale of which varies from $\delta_\nu$ to $\delta$. From a physical viewpoint, this is consistent with the underlying rationale in the derivation of (\ref{eq:1.1}). In this section, a further progress from (\ref{eq:2.5}) or alternatively from (\ref{eq:2.6}) will be sought by considering two scenarios associated with the $Re_\tau$-dependent behaviour of the spectrum in Region I. In particular, it will be shown that the two scenarios establish theoretical links with the scaling laws in (\ref{eq:1.1}) \cite[]{Marusic2003} and (\ref{eq:1.3}) \cite[]{Monkewitz2015}, respectively.

\subsection{Scenario 1: $h_1(k_x \delta;Re_\tau)>0$ for all $Re_\tau$}\label{sec:3.1}
The first scenario assumes that the pre-multiplied spectrum in Region I, $h_1(k_x \delta; Re_\tau)$, stops to decay to zero above a certain $Re_\tau$, and it eventually loses the $Re_\tau$ dependence, making $A(Re_\tau)$ non-zero constant as $Re_\tau \rightarrow \infty$ (see also (\ref{eq:2.5b})). This scenario is not yet supported by the DNS and experimental data in figure \ref{fig3}, as they do not show such a behaviour up to $Re_\tau = 20000$. However, here I shall assume that this may happen for very high $Re_\tau$ for the purpose of understanding its theoretical consequences. In particular, this will lead to the result directly connected to the scaling law of (\ref{eq:1.1}).

\begin{theorem}\label{st:1}
\textit{If $h_1(k_x \delta; Re_\tau) \rightarrow \tilde{h}_1(k_x \delta)>0$ as $Re_\tau \rightarrow \infty$,
\begin{equation}
\frac{\overline{u'u'}}{u_\tau^2} \rightarrow \infty\quad at \quad y^+=c
\end{equation}
with a lower bound proportional to $\ln Re_\tau$.}
\end{theorem}
Since $Re_\tau$ dependence of the spectrum originates from Region I, the given assumption effectively removes $Re_\tau$ dependence of the spectrum not only in Region I but also in Region III: i.e. $h_3(k_x l; Re_\tau) \rightarrow \tilde{h}_3(k_x l)>0$ for $a/\delta \leq k_x \leq b/\delta_\nu$. Note that $a$ can be chosen arbitrarily. Therefore, suppose a sufficiently small $a$ is chosen for all $Re_\tau$, such that $\tilde{h}_1(k_x \delta=a) = D$ where $D \leq \inf_{k_x} \tilde{h}_3(k_x l)$ for $a/\delta \leq k_x \leq b/\delta_\nu$. In this case, the contribution of the spectrum to $\overline{u'u'}/u_\tau$ in Region III must satisfy the following inequality (see also figure \ref{fig4} for a schematic sketch):
\begin{equation}\label{eq:3.1}
\int_{\ln (al/\delta)}^{\ln (bl/\delta_\nu)} \tilde{h}_3(k_x l) \mathrm{d}(\ln (k_x l)) \geq D \ln Re_\tau +\ln \left(\frac{b}{a}\right).
\end{equation}
This indicates that $\overline{u'u'}/u_\tau$ is unbounded as $Re_\tau \rightarrow \infty$ and it has a lower bound proportional to $\ln Re_\tau$ as stated above.

A stronger result can further be obtained, if the result of the classical theory of \cite{Perry1986} is combined.
\begin{remark}\label{st:2}
\textit{If $h_1(k_x \delta; Re_\tau) \rightarrow \tilde{h}_1(k_x \delta)>0$ as $Re_\tau \rightarrow \infty$,
\begin{equation}
\frac{\overline{u'u'}}{u_\tau^2} \sim \ln Re_{\tau} \quad at \quad y^+=c.
\end{equation}
}
\end{remark}
If $h_1(k_x \delta; Re_\tau) \rightarrow \tilde{h}_1(k_x \delta)(>0)$, one can choose $a$ so that $A(Re_\tau)=A_0$ from (\ref{eq:2.5b}), where $A_0$ is a positive non-zero constant. Importantly, in this case, the entire theoretical setting here becomes identical to that of \cite{Perry1986}, if their wall-normal coordinate is replaced with $Re_\tau$. Following the result of \cite{Perry1986}, the asymptotic form of the pre-multiplied spectrum in Region III for $1/\delta \ll k_x\ll b/\delta_\nu$ is then given by the wall-known $k_x^{-1}$ spectrum: i.e.
\begin{equation}
\frac{k_x\Phi_{uu}(k_x)}{u_\tau^2}=h_3(k_x l; Re_\tau) = \tilde{h}_3(k_x l)={B_0},
\end{equation}
where $B_0$ is a positive non-zero constant. Since $A(Re_\tau)=A_0$ and $B(Re_\tau) \simeq B_0$ from (\ref{eq:2.5d}), this finally yields
\begin{equation}\label{eq:3.3}
\frac{\overline{u'u'}}{u_\tau^2}\Big|_{y^+=c} \simeq A_0+C+B_0\left[\ln Re_\tau+\ln \left(\frac{b}{a}\right)\right].
\end{equation}

Now, it is evident that the consequence of assuming non-Reynolds-number dependence of the spectrum in Region I yields the scaling in (\ref{eq:1.1}): comparing (\ref{eq:1.1}) with (\ref{eq:3.3}) leads to $B_s=B_0$ and $A_s=A_0+B_0+C \ln (b/a)$. Furthermore, $C$ in (\ref{eq:3.3}) (i.e. contribution from Region II in figure \ref{fig4}) is equivalent to the contribution of $f_1$ in (\ref{eq:1.2a}) and the rest in (\ref{eq:3.3}) corresponds to the contribution of $f_2$ in (\ref{eq:1.2a}) in the form of (\ref{eq:1.2b}). It is important to note that removing the $Re_\tau$-dependence of the spectrum in Region I is identical to ignoring the viscosity in Region I and III, resulting in an inviscid assumption for the motions associated with the logarithmic and outer regions, as in the theory based on attached eddy hypothesis \cite[]{Marusic2003}. 

\subsection{Scenario 2: $\overline{u'u'}/u_\tau^2 < \infty$ as $Re_\tau \rightarrow \infty$}\label{sec:3.2}
Although Scenario 1 in \S\ref{sec:3.1} gives a result identical to the classical theory, the DNS and experimental data in figure \ref{fig3} do not appear to strongly support it. Indeed, the pre-mulitplied spectrum in Region I appears to decay slowly on increasing $Re_\tau$, and it does not show any $k_x^{-1}$ behavior in Region III. This is more evident, given that the spectra from DNS and experimental data (figure \ref{fig3}) rather clearly show a $Re_\tau$-independence in Region II. It rather deems possible that the pre-mulitplied spectrum reaches zero as $Re_\tau \rightarrow 0$: i.e. $h_1(k_x \delta; Re_\tau) \rightarrow 0$. However, this condition alone does not provide a further insight. This is because the scaling of $\overline{u'u'}/u_\tau^2$ essentially depends on how quickly the unknown pre-multiplied spectrum $h_3(k_x l; Re_\tau)$ in Region III decays: see (\ref{eq:2.5}). A stronger assumption would therefore be needed in order to make a further theoretical progress. Fortunately, \cite{Monkewitz2015} previously showed that is is possible to have $\overline{u'u'}/u_\tau < \infty$ in boundary layers, as $Re_\tau \rightarrow \infty$. This condition provides a useful guideline on how quickly $h_3(k_x l; Re_\tau)$ in Region III would have to decay. If this condition is employed, the following result is obtained.

\begin{theorem}\label{st:3}
\textit{Suppose $B(\sigma)\in \mathbb{C}^\infty$ at $\sigma=0$ with $\sigma=1/\ln Re_\tau$. If $\overline{u'u'}/u_\tau < \infty$ as $Re_\tau \rightarrow \infty$ and either $\mathrm{d}B/\mathrm{d}\sigma$ or $\mathrm{d}^2 B/\mathrm{d}\sigma^2$ is non-zero at $\sigma=0$,
\begin{equation}
\frac{\overline{u'u'}}{u_\tau^2} \sim \frac{1}{\ln Re_{\tau}} \quad at \quad y^+=c.
\end{equation}
}
\end{theorem}

From (\ref{eq:2.6a}), if $\overline{u'u'}/u_\tau^2 < \infty$ is to be satisfied as $Re_\tau \rightarrow \infty$, it must be that $B(Re_\tau) \rightarrow 1/(\ln Re_\tau)^n$ with $n\geq 1$. Therefore, $1/\ln Re_\tau$ now naturally emerges as a physically relevant small parameter for $Re_\tau \rightarrow \infty$. Since $B(\sigma)\in \mathbb{C}^\infty$ is assumed at $\sigma=0$, taking $\sigma$ as the small parameter of interest yields
\begin{equation}\label{eq:3.4}
B(\sigma)=B'(0)\sigma+\frac{B''(0)}{2}\sigma^2+O(\sigma^3),
\end{equation}
where $(\cdot)'$ for $B$ indicates differentiation with respect to $\sigma$. As $B'(\sigma)$ and $B''(\sigma)$ were assumed to be non-zero, this ultimately leads to the following scaling law:
\begin{subequations}\label{eq:3.5}
\begin{equation}\label{eq:3.5a}
\frac{\overline{u'u'}}{u_\tau^2}\Big|_{y^+=c}=E+\frac{F}{\ln Re_\tau}+O(\sigma^2),
\end{equation}
where
\begin{equation}
E=C+B'(0),
\end{equation}
\begin{equation}
F=\frac{B''(0)}{2}+B'(0)\ln\left(\frac{b}{a}\right).
\end{equation}
\end{subequations}
Note that the assumption of $B'(0) \ne 0$ here is equivalent to assuming that $B(Re_\tau) \rightarrow 0$ as $Re_\tau \rightarrow \infty$ with the slowest possible rate to satisfy $\overline{u'u'}/u_\tau^2 < \infty$: i.e. $B(Re_\tau) \rightarrow 1/\ln Re_\tau$. In this case, it must be that $B'(0)>0$, because the $B(Re_\tau)>0$ for all $Re_\tau$ by definition. Since $E$ is expected be the least upper bound of $\overline{u'u'}/u_\tau^2$ as $Re_\tau \rightarrow \infty$, this implies $F<0$, resulting in the following condition for $B''(0)$ to satisfy:
\begin{equation}\label{eq:3.6}
B''(0)<-B'(0)\ln \left(\frac{b}{a}\right).
\end{equation}
Furthermore, since $U_\infty^+ \sim \ln Re_\tau$ for $Re_\tau \rightarrow \infty$, (\ref{eq:3.5a}) is asymptotically equivalent to the result from \cite{Monkewitz2015} given in (\ref{eq:1.3}). However, unlike their analysis applicable only to boundary layers, the $1/\ln Re_\tau$ scaling derived here is based on the structure of near-wall spectrum of streamwise velocity. Therefore, it is applicable to other flows, such as channel and pipe. Importantly, the framework introduced here unifies (\ref{eq:1.1}) from the classical attached eddy model with (\ref{eq:1.3}) from an asymptotic analysis. The former is a consequence of not accounting for the viscous effect of the motions from the log and outer regions. The latter is the scaling law obtained by assuming the outer-scaling part of the spectrum decays with the slowest possible rate to satisfy $\overline{u'u'}/u_\tau^2 < \infty$ for all $Re_\tau$.

\begin{figure} \vspace*{2mm}
\centering
\includegraphics[width=0.95\textwidth]{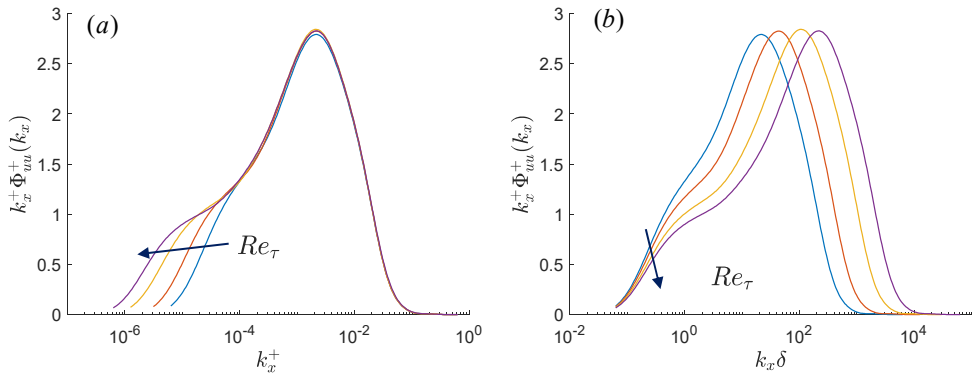}
\caption{Premultiplied streamwise wavenumber spectra of streamwise velocity at $y^+=20$ in channel \cite[from a quasilinear approximation by][]{Holford2023b}: $(a)$ inner scaling; $(b)$ outer scaling. Here, .}\label{fig5}
\end{figure}

Finally, given the formulation introduced here, (\ref{eq:3.5}) is supposed to be valid strictly at a fixed inner-scaling wall-normal location: i.e. $y^+=c$. However, the slow change of the peak wall-normal location with $Re_\tau$ \cite[]{Willert2017,Hwang2020a} can further be taken into account. Taking $\sigma(=1/\ln Re_\tau)$ as the small parameter of physical relevance, the peak streamwise turbulence intensity is written further as
\begin{subequations}
\begin{equation}
\frac{\overline{u'u'}}{u_\tau^2}\Big|_{y^+=y^+_p}=\frac{\overline{u'u'}}{u_\tau^2}\Big|_{y^+=c}+\frac{G}{\ln Re_\tau} +O(\sigma^2),
\end{equation}
where
\begin{equation}
G=\frac{1}{u_\tau^2}\frac{\partial \overline{u'u'}}{\partial y^+}\Big|_{y^+=c}\frac{\mathrm{d} y^+}{\mathrm{d} \sigma}\Big|_{\sigma=0}.
\end{equation}
\end{subequations}
Combining with (\ref{eq:3.5a}), this ultimately gives
\begin{equation}\label{eq:3.7}
\frac{\overline{u'u'}}{u_\tau^2}\Big|_{y^+=y^+_p}=E+\frac{H}{\ln Re_\tau}+O(\sigma^2),
\end{equation}
where $H=F+G$, and it still retains in the form of (\ref{eq:3.5a}).

\section{Validation}\label{sec:4}

\begin{figure} \vspace*{2mm}
\centering
\includegraphics[width=0.95\textwidth]{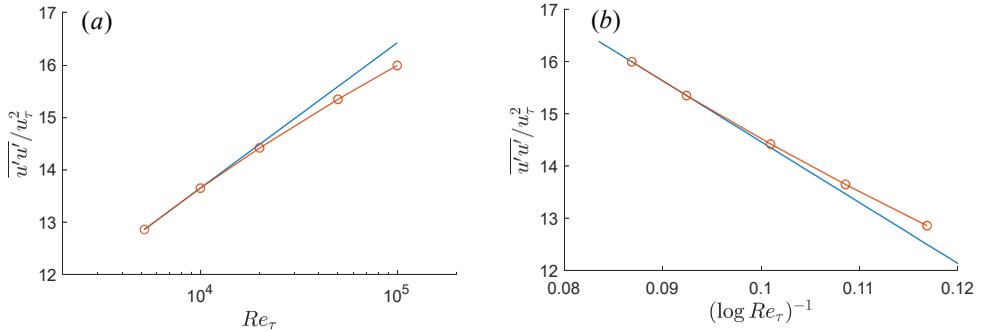}
\caption{Scaling of streamwise turbulence intensity at $y^+=20$ for the quasilinear approximation of \cite{Holford2023b} ($Re_\tau=5200, 10000, 20000, 50000, 100000$): $(a)$ $\ln Re_\tau$ scaling; $(b)$ $1/\ln Re_\tau$ scaling.}\label{fig6}
\end{figure}

In this section, we further justify the theoretical results obtained in \S\ref{sec:3}. However, as shown in figure \ref{fig1}, all the existing data do not provide a clear idea on which of the scalings would fit best. Given that all the asymptotic scalings introduced in \S\ref{sec:1}, it is therefore necessary to have another dataset of physical relevance at higher $Re_\tau$. \cite{Hwang2020a} recently introduced a self-consistent quasilinear approximation of the Navier-Stokes equations for channel flow. In this model, the full nonlinear mean equation is considered, while the fluctuation equations is modelled using the linearised Navier-Stokes equations where the original self-interacting nonlinear term is modelled using an eddy viscosity and stochastic forcing. The quasilinear approximation is computationally very cheap, and has been significantly extended \cite[]{Skouloudis2021,Holford2023b}. In this study, the data from \cite[]{Holford2023b} is further introduced, where streamwise wavenumber spectra and the corresponding turbulence intensity are computed up to $Re_\tau=10^5$. For a further details on the quasilinear model, the reader may refer to \cite[]{Holford2023b}.

Figure \ref{fig5} shows the pre-multiplied spectra of streamwise velocity at $y^+=20$ from the quasilinear approximation of \cite[]{Holford2023b}. It is evident that all the key scaling behaviours of the streamwise velocity spectra in with figure \ref{fig4} are reproduced by the quasilinear approximation: for $k_x^+ \geq 10^{-4}$, the spectra remain universal, while for $k_x \delta \sim O(1)$, they show a decaying behaviour with $Re_\tau$, consistent with those from DNS and experiment in figure \ref{fig3}. The difference between the spectra from the quaslinear model and DNS/experiment is merely quantitative: for example, the peak value of the spectra from DNS and experiment is about $2.2$, while that from the quasilinear approximation is about $2.8$. However, as seen in the theoretical analysis in \S\ref{sec:3}, this is not important for the purpose of discussing the relevant scaling of data.

In figure \ref{fig6}, the scaling of the near-wall streamwise turbulence intensity from the quasilinear approximation at $y^+=20$ is reported. The intensity appears to follow a $\ln Re_\tau$ scaling at least approximately up to $Re_\tau \simeq 2 \times 10^4$, consistent with the experimental data including \cite{Samie2018} (figure \ref{fig6}$a$). However, similarly to the observation of \cite{Willert2017} from the CICLoPE facility of the University of Bologna, it is seen that the intensity begins to deviate from the $\ln Re_\tau$ scaling from $Re_\tau \simeq 2 \times 10^4$ and this deviation becomes gradually larger, clearly visible at $Re_\tau = 1 \times 10^5$. Interestingly, it appears that the same data begin to follow an $1/\ln Re_\tau$ scaling for $Re_\tau \geq 2 \times 10^4$ (figure \ref{fig6}$b$). Note that exactly the same behaviour has been observed from a computational more economical variant of the quasilinear approximation \cite{Skouloudis2021}, where the near-wall streamwise turbulence intensity is obtained up to $Re_\tau = 1 \times 10^6$.

\begin{figure} \vspace*{2mm}
\centering
\includegraphics[width=0.95\textwidth]{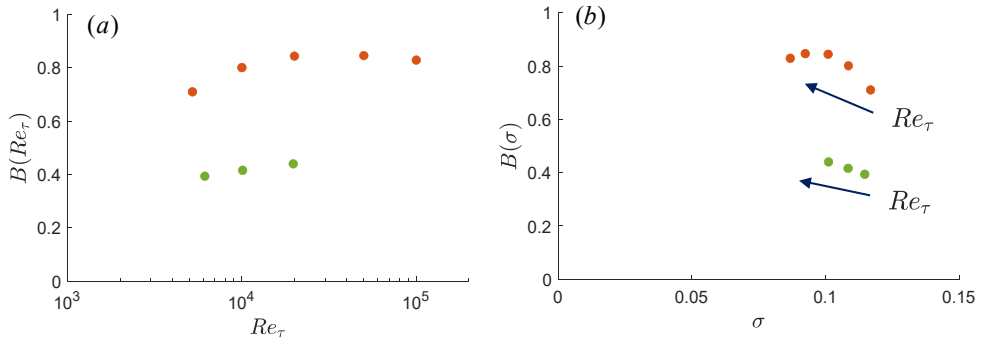}
\caption{$Re_\tau$ dependence of $B(Re_\tau)$ in (\ref{eq:2.5a}): $(a)$ $B(Re_\tau)$ vs $\sigma(=(1/\ln Re_\tau)^{-1})$; $(b)$ $B(\sigma)$ vs $Re_\tau$. Here, data are from quasilinear approximation of \cite{Holford2023a} (orange) with $a=0.1$ and $b\simeq 1 \times 10^{-4}$ and experiment of \cite{Samie2018} (green) with $a=0.2$ and $b\simeq 3 \times 10^{-4}$. Here, $Re_\tau=5200, 10000, 20000, 50000, 100000$ for the quasilinear approximation and $Re_\tau=6123, 10100, 19680$ for the experiment.}\label{fig7}
\end{figure}

The data shown in figures \ref{fig1} and \ref{fig6} suggest that there may be a transition in the behaviour of $\overline{u'u'}/u_\tau^2$ at $Re_\tau \sim O(10^4)$. To understand this behaviour better, $B(Re_\tau)$ (or $B(\sigma)$ with $\sigma=1/\ln(Re_\tau)$) in (\ref{eq:2.6a}) is further calculated using the velocity spectra from both the quasilinear approximation and the experiment of \cite{Samie2018}. Note that $B(Re_\tau)$ is the Reynolds-number dependent part of the turbulence intensity obtained in (\ref{eq:2.6a}), when a sufficiently small value of $a$ is chosen. Figure $\ref{fig7}$ shows $B(Re_\tau)$ with respect to $Re_\tau$ and $\sigma$. Both quasilinear approximation and experimental data show that $B(Re_\tau)$ slowly grows with $Re_\tau$ up to $Re_\tau \simeq 2 \times 10^4$ (figure \ref{fig7}$a$). It is unfortunate that the experimental data above $Re_\tau \simeq 2 \times 10^4$ is not available, but in $B(Re_\tau)$ from the quasilinear approximation stops growing around $Re_\tau \simeq 2 \times 10^4$. For $Re_\tau \geq 5 \times 10^4$, the $B(Re_\tau)$ begins to decay. This behaviour is more clearly seen when $B(\sigma)$ is plotted (figure \ref{fig7}$b$). It is worth reminding that if $\overline{u'u'}/u_\tau^2 < \infty$ as $Re_\tau \rightarrow \infty$, it must be that $B(\sigma) \rightarrow 0$ as $\sigma \rightarrow 0$. The $B(\sigma)$ from the quasilinear approximation begins to exhibit this behavior around $Re_\tau \simeq 2 \times 10^4$, where $\overline{u'u'}/u_\tau^2$ is approximately deviated from the $\ln Re_\tau$ scaling. Furthermore, the form of $B(\sigma)$ around $Re_\tau \simeq 2 \times 10^4$ suggests that it is possible to have $B''(0)<0$. Since $\ln (b/a)<0$ (from the choice of $a(=0.1)$ and $b\simeq 10^{-4}$ in figure \ref{fig7}) and $B'(0)$ is likely to be positive from the data, this also satisfies the condition in (\ref{eq:3.6}).

\begin{figure} \vspace*{2mm}
\centering
\includegraphics[width=0.75\textwidth]{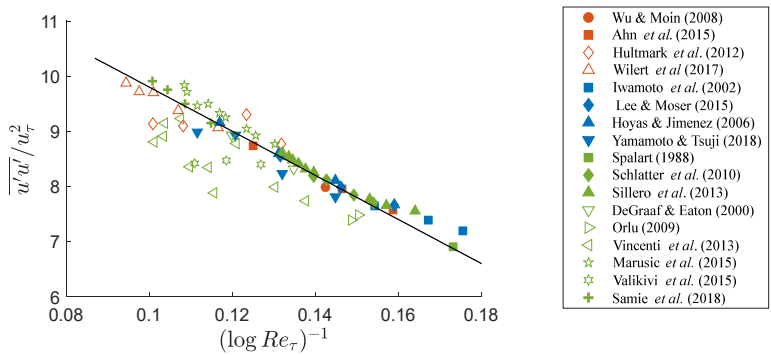}
\caption{Peak streamwise turbulence intensities from experiments (open symbols) and DNSs (filled symbols) with the $1/\ln Re_\tau$ scaling.}\label{fig8}
\end{figure}

The examination of the quasilinear approximation data suggests that the $1/\ln Re_\tau$ scaling discussed in \S\ref{sec:3.2} is a certainly plausible scenario. However, it is seen that the existing experimental data do not appear to cover the regime of $Re_\tau$, in which the possible transition in the behaviour of $B(Re_\tau)$ might take place. That being said, it is important to note that the existing experimental data do not exclude the classical $\ln Re_\tau$ scaling yet. Unlike the quasilinear approximation data, it is still possible that the outer-scaling part of the pre-multiplied spectrum, $h_1(k_x \delta; Re_\tau)$, may be converged to non-zero values around a certain $Re_\tau > O(10^4)$. In this case, the classical attached model is truly applicable as demonstrated in \S\ref{sec:3.1}, resulting in an indefinite logarithmic growth of turbulence intensity as $Re_\tau \rightarrow$. From this viewpoint, high fidelity experimental or numerical simulation data for $Re>O(10^4)$ would still be required to resolve this on-going debate, although the overall trend of the existing data appears to favour the $1/\ln Re_\tau$ scaling while excluding the possibility of the scaling proposed by \cite{Chen2021}. Given the discussion above, it would finally be worth plotting all the DNS and experimental data with respect to $1/\ln (Re_\tau)$, and this is shown in figure \ref{fig8}. The collapse of data with a linear fit of $1/\ln (Re_\tau)$ is as good as that with the $\ln (Re_\tau)$ scaling shown in figure \ref{fig1}, indicating the difficulty in identifying the correct asymptotic scaling behaviour using the existing data.

\section{Concluding remarks}\label{sec:5}
In this study, the asymptotic behaviour of $\overline{u'u'}/u_\tau^2$ has been explored. The three scalings previously proposed \cite[]{Marusic2003,Monkewitz2015,Chen2021} have been reviewed with their derivation process and physical justification. A new analysis has subsequently been introduced based on velocity spectrum, and two possible scenarios have been identified. Scenario 1 assumes that the outer-scaling part of the near-wall velocity spectrum reaches a non-zero constant as $Re_\tau \rightarrow \infty$. The resulting scaling law was the classical $\ln Re_\tau$ scaling of \cite{Marusic2003}, physically consistent with the classical attached eddy model \cite[]{Townsend1976,Perry1986}. Scenario 2 assumes a sufficiently rapid decay of the outer-scaling part of the near-wall velocity spectrum with $Re_\tau$ due to the effect of viscosity, such that $\overline{u'u'}/u_\tau^2 < \infty$ for all $Re_\tau$. In this case, an $1/\ln Re_\tau$ law has been obtained, and this is consistent with the result of the asymptotic analysis for the streamwise mean momentum equation in a boundary layer \cite[]{Monkewitz2015}. The two scenarios have been further verified with data from a quasilinear approximation \cite{Holford2023b}, which favour Scenario 2. However, to extend this conclusion to real flows, high-fidelity new measurements for $Re_\tau \gtrsim O(10^4)$ would be necessary.

Finally, it is worth making some remarks on the $Re_\tau^{-1/4}$ scaling by \cite{Chen2021}. As discussed in \S\ref{sec:1.3}, the scaling is based on two questionable conjectures. As such, there is no strong physical reason that $Re_\tau^{-1/4}$ must be the relevant scaling parameter for near-wall streamwise turbulence intensity as $Re_\tau \rightarrow \infty$. A similar issue has also been raised by \cite{Monkewitz2022}, who pointed out that the $Re_\tau^{-1/4}$ scaling does not naturally emerge in the Reynolds stress transport equation for $\overline{u'u'}/u_\tau^2$, whereas $1/\ln(Re_\tau)$ appears naturally from the velocity spectrum and from the streamwise mean momentum equation in boundary layer \cite[]{Monkewitz2015}. In any case, if $\overline{u'u'}/u_\tau^2 \sim Re_\tau^{-1/4}$ is assumed, $B(Re_\tau)=Re_\tau^{-1/4}/\ln(Re_\tau)$ can be set from (\ref{eq:2.6a}). This implies $B(\sigma)=\sigma e^{-1/(4\sigma)}$ as $\sigma \rightarrow 0$. Since $e^{-1/4\sigma}$ shall decay zero much more slowly than $\sigma$ as $\sigma \rightarrow 0$, it would be difficult to find any numerical difference between $B(\sigma) \sim \sigma$ ($1/\ln Re_\tau$ scaling) and $B(\sigma)=\sigma e^{-1/(4\sigma)}$ ($Re_\tau^{-1/4}$ scaling) in practice. This was also recently pointed out by \cite{Monkewitz2022} and by \cite{Nagib2022}.


\section*{Acknowledgement}
This work was initiated when I was visiting the Isaac Newton Institute for Mathematical Sciences at the University of Cambridge for the
programme `Mathematical aspects of turbulence: where do we stand?'. I would like to thank Prof. Rich Kerswell, who encouraged me to start this work during the visit, and Prof. I. Marusic, who shared the initial discussion of the paper. I would also like to thank Prof. Peter Monkewitz and Prof. Hasan Nagib, who made me realise this issue and shared an interesting discussion.

\section*{Funding}
This work was supported by the financial support of the Engineering and Physical Sciences Research Council (EPSRC; EP/T009365/1) in the UK. My visit to the Isaac Newton Institute for Mathematical Sciences at the University of Cambridge was supported by EPSRC through the programme `Mathematical aspects of turbulence: where do we stand?' (EP/R014604/1).

\section*{Declaration of interest}
The authors report no conflict of interest.

\bibliographystyle{jfm}
\bibliography{references}

\end{document}